\documentclass[aps,prb,twocolumn,showpacs,showkeys,groupedaddress,floatfix,amsmath,amssymb]{revtex4}
\usepackage{graphicx}
\usepackage{dcolumn}
\usepackage{bm}

\begin{document}



\title{Masking effect of heat dissipation on the current-voltage characteristics of
a mesoscopic superconducting square with leads}

\author{D.Y. Vodolazov$^{1,2}$}
\author{F.M. Peeters$^1$}
\email{francois.peeters@ua.ac.be} \affiliation{$^1$Departement Fysica, Universiteit Antwerpen (Campus Drie Eiken), B-2610 Antwerpen,
Belgium \\
$^2$ Institute for Physics of Microstructures, Russian Academy of Sciences, 603950, Nizhny Novgorod, GSP-105, Russia}
\author{M. Morelle}
\author{V.V. Moshchalkov}
\affiliation{Nanoscale Superconductivity and Magnetism Group, Laboratory for Solid State Physics and Magnetism, K.U. Leuven, Celestijnenlaan 200D, B-3001
Leuven, Belgium}
\date{\today}

\date{\today}

\begin{abstract}

A theoretical analysis based on a numerical solution of the coupled time-dependent Ginzburg-Landau and heat dissipation equations shows a strong
dependence of the critical currents on the applied magnetic field in a mesoscopic square with attached contacts. In agreement with experiment we found
hysteresis which are caused by a strong heat dissipation in the sample at currents close to the depairing Ginzburg-Landau current and/or the dynamics of
the superconducting condensate. The theoretically obtained nonmonotonous dependence of the switching current (from superconducting to the resistive
state) on the applied magnetic field, arising from the changes in the vorticity, agrees quantitatively with the experimental data. Our results show that
heat dissipation leads to an increase of the hysteresis in the current-voltage characteristic and hence masks the actual dynamics of the superconducting
condensate.

\end{abstract}

\pacs{74.25.Op, 74.20.De, 73.23.-b}

\maketitle

\section{Introduction}

After the discovery of the step-like features in the current-voltage characteristics of superconducting whiskers \cite{Meyer} and the explanation of this
effect through the nucleation of phase slip centers \cite{Skocpol,Dolan} this phenomenon attracted a lot of attention from several experimental and
theoretical groups (see Refs.\cite{Ivlev,Tidecks} for reviews). This effect is a consequence of the nontrivial dynamics of the superconducting condensate
and the normal quasiparticles near the phase slip center \cite{Skocpol,Ivlev,Tidecks}. During the phase slip process the gap goes periodically in time to
zero in one point along the sample and it creates an excess of quasiparticles (electron-like and hole like) near this region. Due to the relatively large
time of relaxation of nonequilibrium quasiparticles in the superconductor they may diffuse on a distance much larger than the coherence length (the size
of suppression of the gap) \cite{Thinkham}. It results in the existence of a nonzero electrical field (as a response to the gradient of chemical
potential of nonequilibrium quasiparticles) and a finite time-dependent voltage drop near the phase slip center. Unfortunately, strong heat dissipation
masks this effect at temperatures far from the critical temperature \cite{Skocpol} T$_c$ which prohibited the study of this effect in full details.

Recently, this subject was revisited because new experimental techniques were developed which made it possible to prepare samples with low resistance
\cite{Michotte} (i.e. diminishing the heating effects at low temperatures) or by using pulsed techniques \cite{Jelila, Michotte2}. The existence of phase
slip centers/lines was confirmed in high-temperature superconductors \cite{Jelila} and they were found to lead to S-shaped I-V characteristics in the
voltage driven regime \cite{Vodolazov1}. Recently phase slip lines were directly observed \cite{Ustinov} in low temperature superconducting stripes. In
Refs. \cite{Kurin,Vodolazov2} a new type of vortex dynamics (so called vortex channelling \cite{Vodolazov2} or the appearance of "kinematical" vortices
\cite{Kurin}) was proposed which, in our opinion, is the "bridge" between slow vortex motion and the fast phase slip line regime. Furthermore, in
Ref.\cite{Bezryadin} the experimental observation of quantum phase slips was claimed.
\begin{figure}[hbtp]
\includegraphics[width=0.35\textwidth]{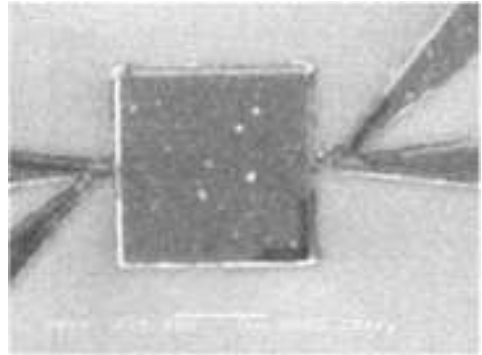}
\caption{SEM micrograph of an Al square with lateral dimension of 2~$\mu$m.}\label{Fig:SEMSquare}
\end{figure}

In this paper we present experimental results on the current-voltage characteristics of a mesoscopic superconducting square with leads. In this geometry,
the contacts play a crucial role (see Fig. 1) because the current density is maximal in them. Therefore we cannot consider those contacts in equilibrium.
This is essentially different from previous work on superconducting film/bridge attached to "massive" superconducting "banks" where it was possible to
assume the contacts in equilibrium. At low magnetic fields phase slip centers will appear in our sample in the narrowest places, where the current
density is maximal. Because of inevitable heat dissipation the sample can transit locally to the normal state (for weak heat transfer) or to the
superconducting resistive state (for strong heat transfer). In both cases the I-V characteristics are hysteretic due to heat dissipation and/or the
dynamics of the superconducting condensate. Additional complications come from the effect of the magnetic field induced currents in the square on the
phase slip process in the contacts. This makes our system new and to our knowledge this situation was not studied before either experimentally or
theoretically.

\begin{figure*}[hbtp]
\includegraphics[width=0.24\textwidth]{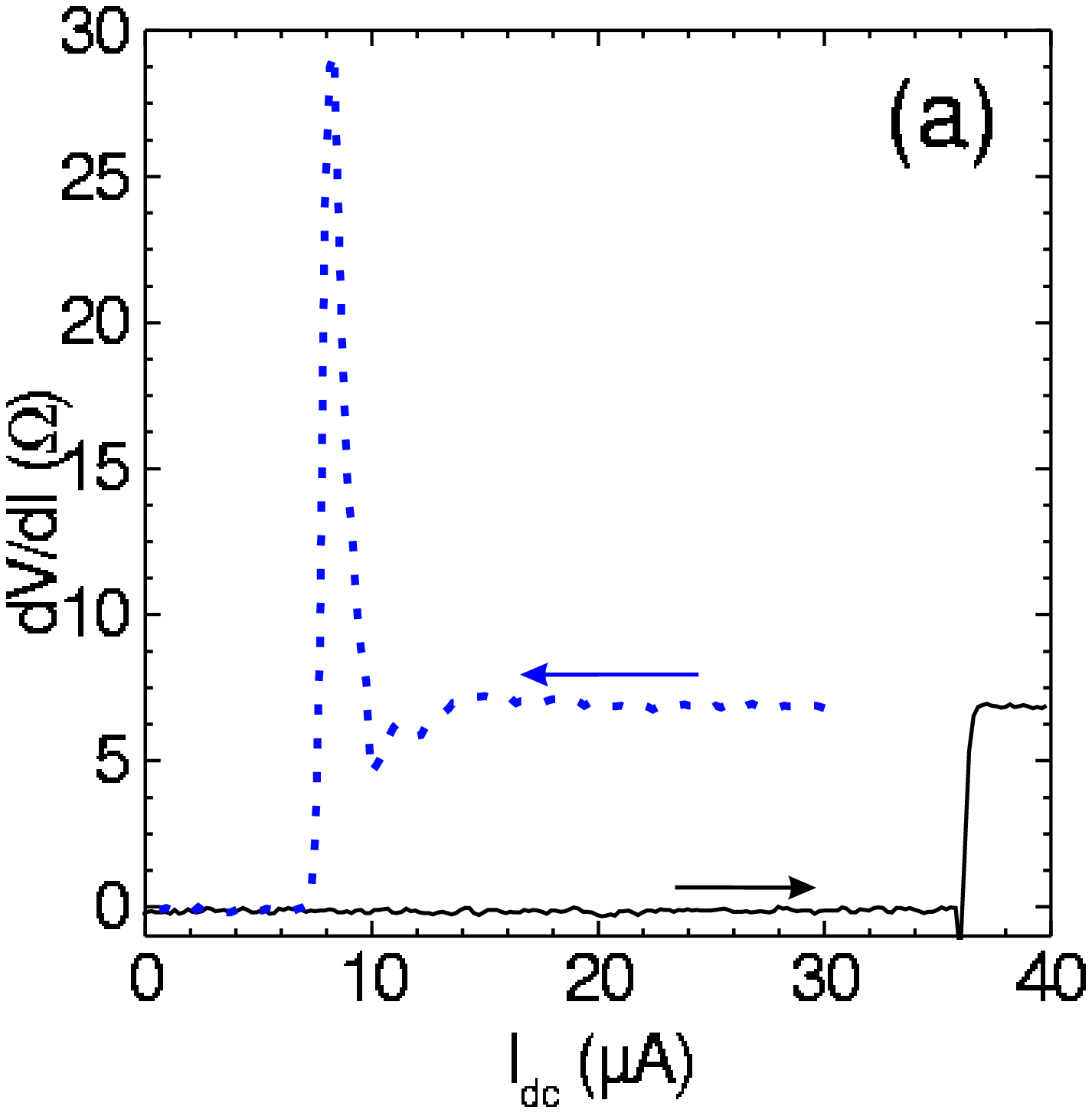}
\includegraphics[width=0.24\textwidth]{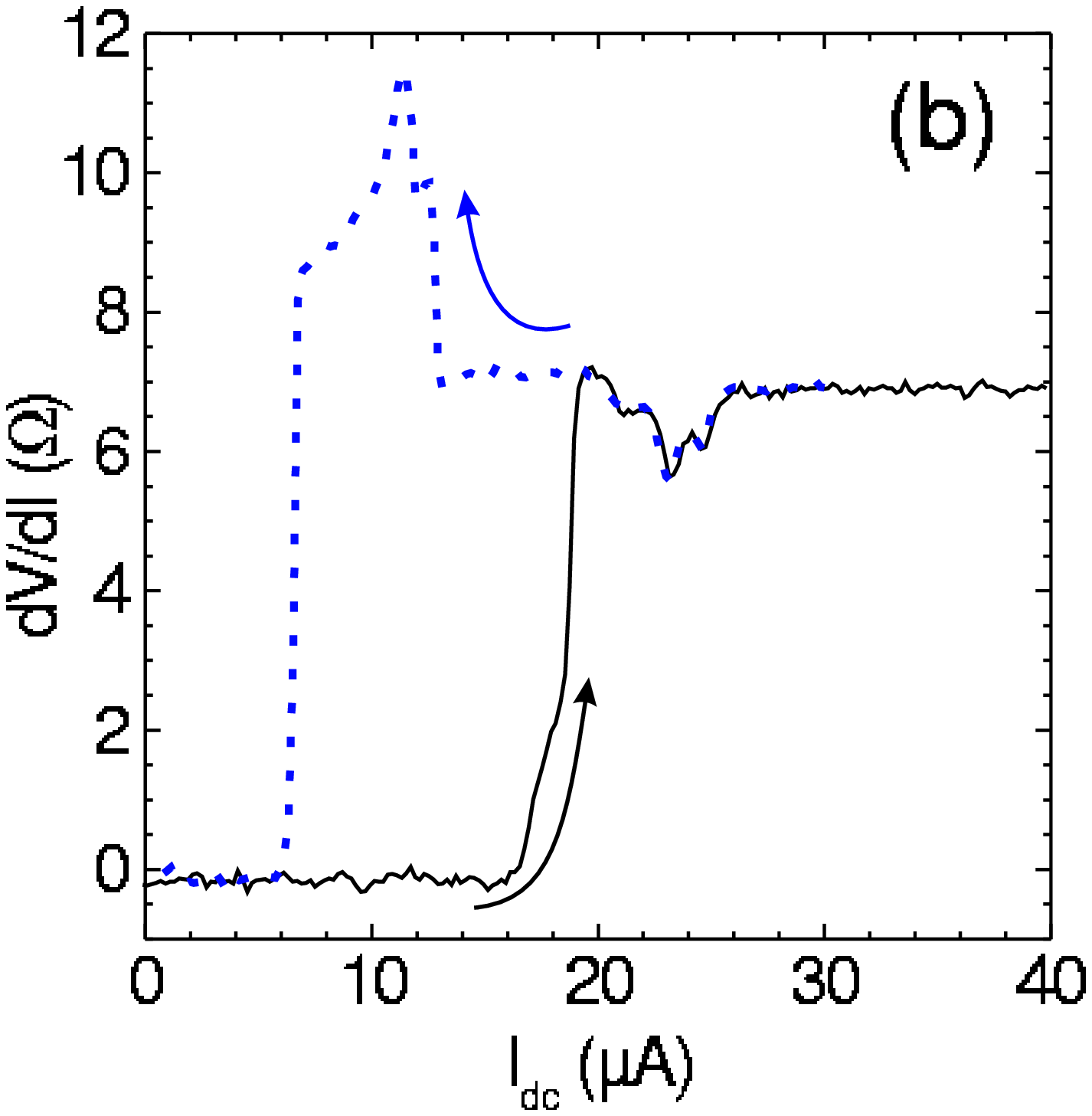}
\includegraphics[width=0.24\textwidth]{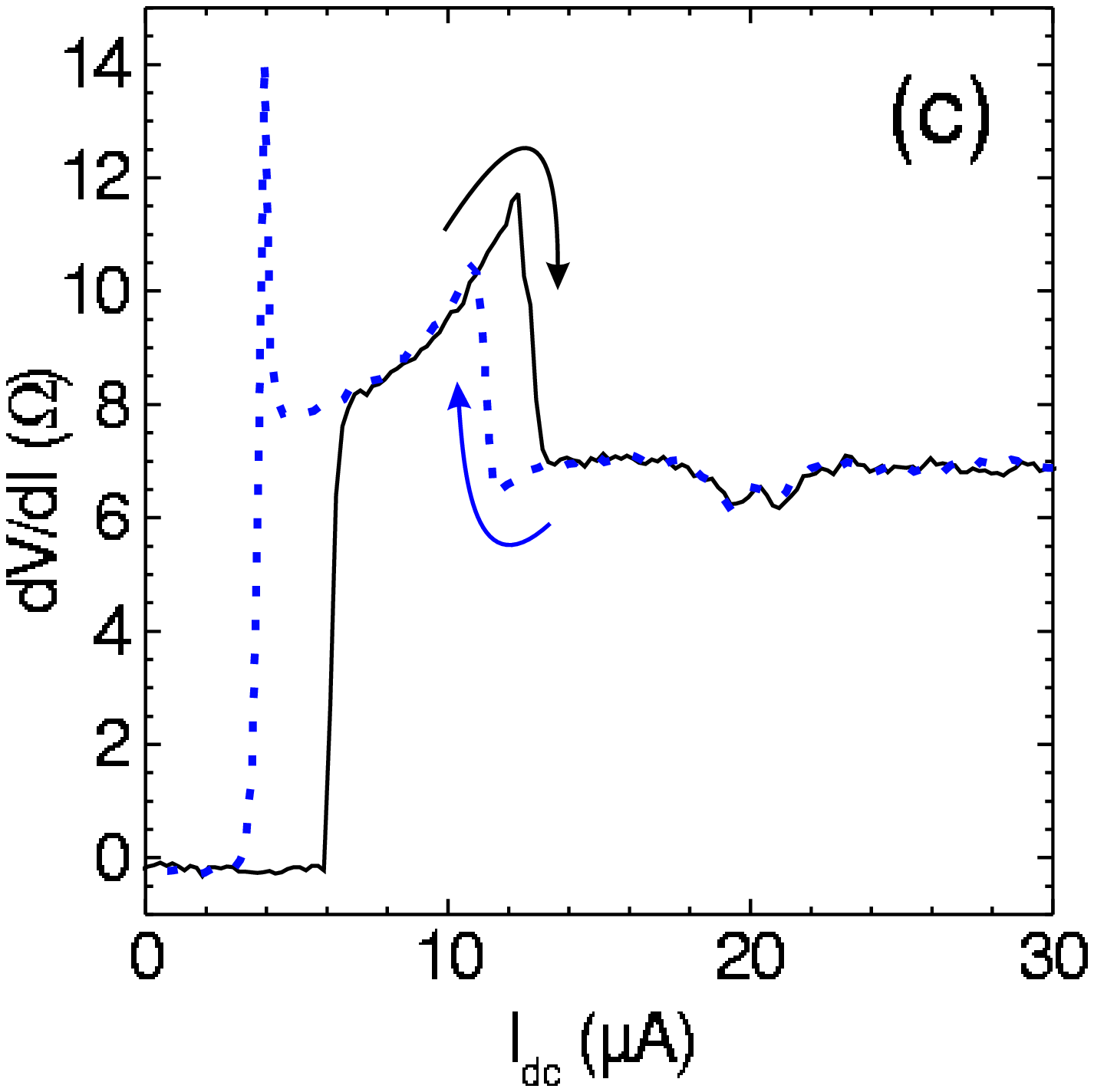}
\includegraphics[width=0.24\textwidth]{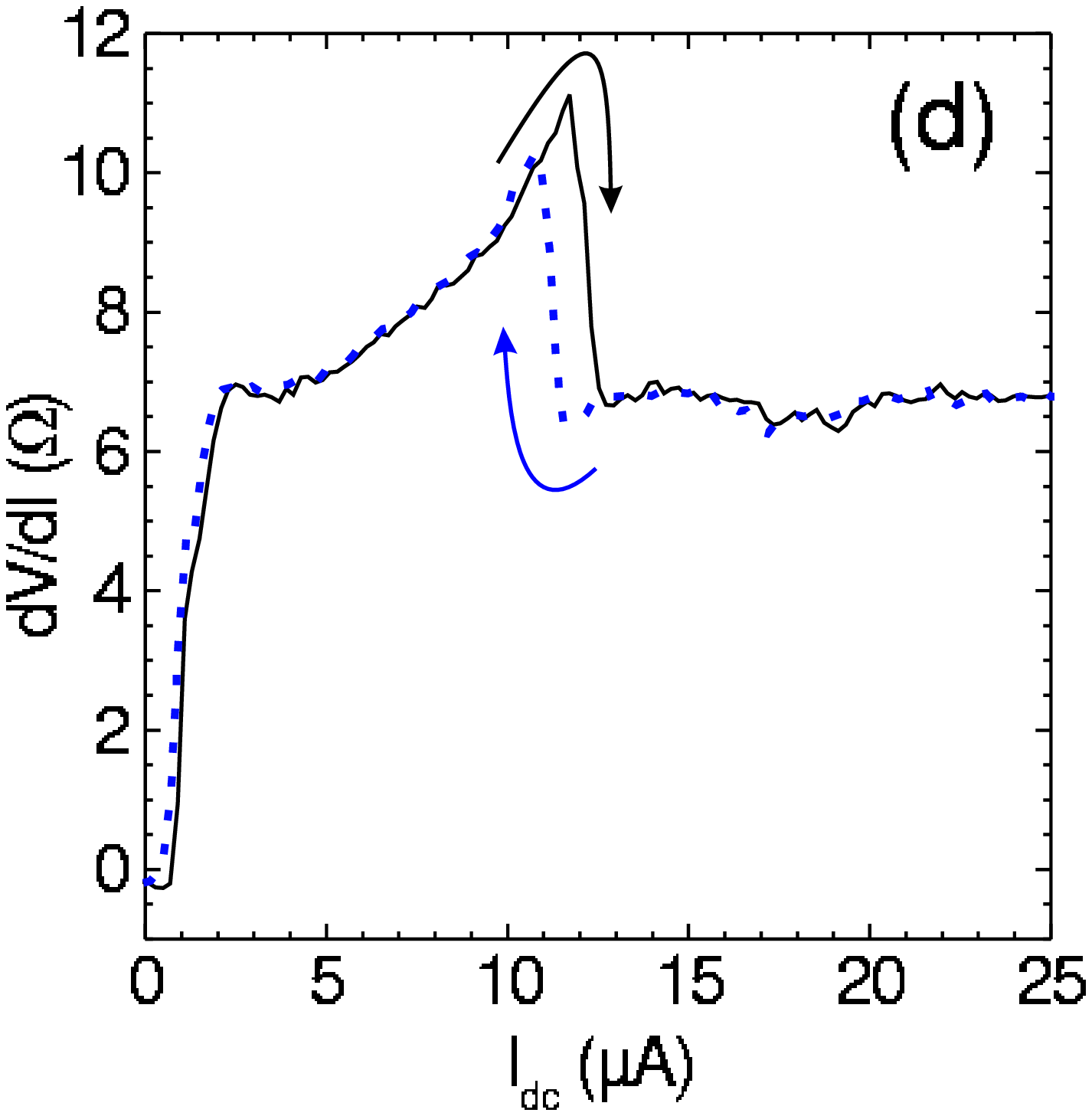}
\caption{ Measured d$V/$d$I$ as a function of the dc current at (a) $\mu_0 H$=0 mT, (b) $\mu_0 H$=3.2 mT, (c) $\mu_0 H$=4.2 mT and (d) $\mu_0 H$=5 mT
measured at $T=1.100$~K. The full black and dashed blue lines are for increasing and decreasing current, respectively.}\label{IVCExp}
\end{figure*}

The paper is organized as follows. In Sec. II we present our experimental results and in Sec. III we give their interpretation on the basis of a solution
of the time-dependent Ginzburg-Landau equations coupled with the equation for heat dissipation. Our conclusions are given in Sec. V.

\section{Experiment}

In Fig.~\ref{Fig:SEMSquare} a SEM micrograph of our superconducting square made from Al using e-beam lithography is shown. The coherence length
determined from a macroscopic co-evaporated sample was found to be $\xi(0)=156$~nm. The thickness was 39 nm found from AFM and from X-ray measurements.
Wedge shaped contacts with an opening angle of $\Gamma=15^\circ$ were used. This shape was used to minimize the effect of the contacts on the
superconducting properties of the square\cite{morelle02,morelle03}.

The experimental $I-V$ characteristics are obtained by superimposing a small ac current (0.1 $\mu$A rms) to a dc current $I_{dc}$. The ac differential
resistance is measured with a EG\&G PAR 124A lock-in amplifier. The dc current is swept from negative to positive value. In order to ensure that the
sample is in the normal state a high dc current is sent through the sample prior to the current sweep. Such sweeps are repeated for different magnetic
fields.

In Fig.~\ref{IVCExp} the differential resistance d$V/$d$I(I_{dc})$ is shown for four values of the magnetic fields. A clear hysteretic behavior is
observed. When starting in the normal state and decreasing the current, the sample remains in a resistive state up to low currents. This resistive state
is not the normal state anymore since a non-constant differential resistance is observed. When starting from the superconducting state and gradually
increasing the current a non-resistive state is observed up to high currents. Contrary to the transition seen at negative currents, a sharp transition
from the non-resistive to the normal state is measured. While in the negative part the transition to the non-resistive state is accompanied by the
appearance of a sharp peak, this is not observed for positive currents. This can be explained by our measuring technique and by the observed hysteretic
behavior. When the transition occurs, the sample remains in the resistive state even when decreasing slightly the current so that the measured ac voltage
will either reflect the non-resistive or the resistive state, but not the transition. Above 1.2 mT, when the first vortex enters the sample, the
transition to the resistive state is preceded by a small increase of the differential resistance [see Fig.~\ref{IVCExp}(b)]. The shape of this part
strongly depends on the vorticity of the sample suggesting a dissipation caused by vortex motion. At high magnetic fields, [see arrows in
Fig.~\ref{IVCExp}(b)]  small symmetric features are observed at high positive and negative currents.

\begin{figure}[hbtp]
\includegraphics[width=0.5\textwidth]{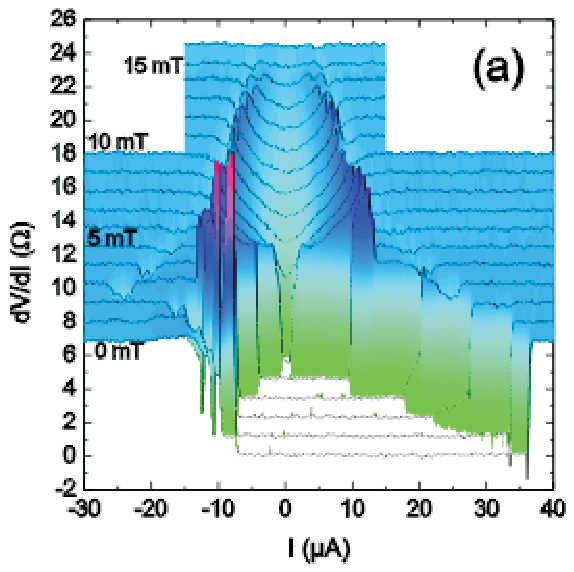}
\includegraphics[width=0.5\textwidth]{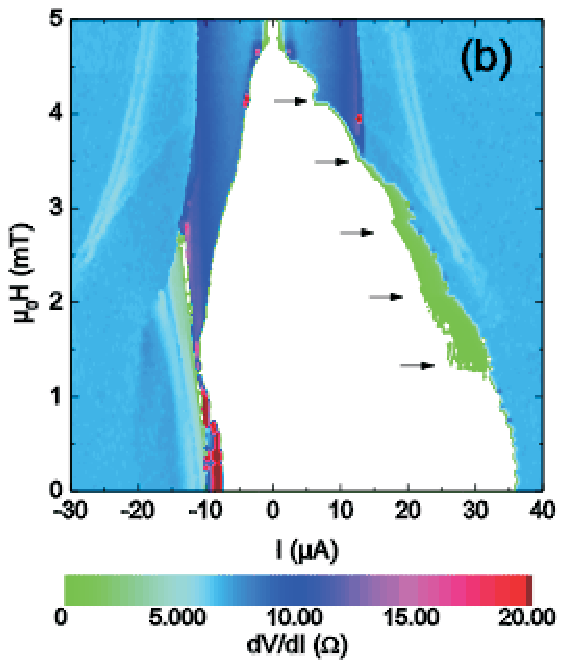}

\caption{(Color online) (a) measured differential resistance d$V/$d$I$ as a function of the dc current for different magnetic fields measured at
$T=1.100$~K. The curves for non-zero field are shifted for clarity. A color map surface is given for comparison with (b). (b) contour plot of the
differential resistance d$V/$d$I$ as a function of dc current and magnetic field measured at $T=1.100$~K for low magnetic fields. The arrows indicate the
points where the vorticity changes.}\label{dVdI}
\end{figure}

In order to study the magnetic field dependence in more details a contour plot of the differential resistance is shown in Fig.~\ref{dVdI} as a function
of the dc current and the magnetic field. The uniform gray (blue) areas at the left and the right correspond to the normal state and the white area in
the middle to the non-resistive state. Oscillations are observed for the transition to the resistive state for positive currents with cusps each time the
vorticity is changed [see arrows in Fig.~\ref{dVdI}(b)]. The dissipation caused by vortex motion, which was already discussed, can also be seen from the
contour plot [light gray (green) area in Fig.~\ref{dVdI}]. The dissipation is only observed for a finite vorticity. At the transition from vorticity
$L$=1 to $L$=2 at $\mu_0H$=2.05 mT, the onset of the dissipation seems to be continuous, but the dissipation increases more rapidly with increasing
current when increasing the vorticity as can also be seen from Fig.~\ref{IVCExp}(b).

Beginning at $\pm 2.5$~mT small symmetric features are seen at high current [white (light blue) lines and the dark gray (dark blue) area in
Figs.~\ref{dVdI}(a)-(b)]. These features are visible up to $\pm 20$~mT and can thus not be arising from the square since the square has a critical field
$H_{c3}^* =6 $~mT. It can only be generated by the contacts which have a higher critical field since a superconducting/normal boundary with a sharp angle
increases the critical field\cite{fomin98epl,schweigert99prbrief}.

Above $\pm 4.5$~mT no hysteretic behavior is seen. All these features are perfectly reproduced on a different sample and are not dependent on the
measuring conditions (sweep rate, amplitude of ac current, temperature). The same behavior is seen when fixing the applied dc current and sweeping the
magnetic field: when the square is in the normal state, the sample remains in a resistive state up to low magnetic fields while when starting from the
non-resistive state and increasing the magnetic field the resistive state is reached only at higher magnetic fields.

\section{Theory}

To understand the experimental results we studied the current-voltage characteristics of two-dimensional superconductors using the generalized
time-dependent Ginzburg-Landau (TDGL) equation \cite{Kramer}
\begin{eqnarray}
\frac{u}{\sqrt{1+\gamma^2|\psi|^2}} \left(\frac {\partial }{\partial t} +i\varphi + \frac{\gamma^2}{2}\frac{\partial|\psi|^2}{\partial t}
\right)\psi= \nonumber \\
=(\nabla - {\rm i} {\bf A})^2 \psi +(1-T-|\psi|^2)\psi.
\end{eqnarray}
where the parameter $\gamma=2\tau_E\Delta_0/\hbar$ is the product of the inelastic collision time $\tau_E$ for electron-phonon scattering and
$\Delta_0=4k_BT_cu^{1/2}/\pi$ is the value of the gap at $T=0$ which follows from Gor'kov's derivation \cite{Gor'kov} of the Ginzburg-Landau equations.

This equation should be supplemented with the equation for the electrostatic potential
\begin{eqnarray} \Delta \varphi & = &
{\rm div}\left({\rm Im}(\psi^*(\nabla-{\rm i}{\bf A})\psi)\right),
\end{eqnarray}
which is nothing else than the condition for the conservation of the total current in the wire, i.e. ${\rm div} {\bf j}=0$. In Eqs. (1,2) all the
physical quantities (order parameter $\psi=|\psi|e^{i\phi}$, vector potential ${\bf A}$ and electrostatical potential $\varphi$) are measured in
dimensionless units: temperature in units of the critical temperature T$_c$, the vector potential ${\bf A}$ and the momentum of the superconducting
condensate ${\bf p}=\nabla \phi -{\bf A}$ are scaled in units $\Phi_0/(2\pi\xi(0))$ (where $\Phi_0$ is the quantum of magnetic flux), the order parameter
in units of $\Delta_0$ and the coordinates are in units of the coherence length $\xi(0)=(8k_BT_c/\pi \hbar D)^{-1/2}$. In these units the magnetic field
is scaled with $H_{c2}=\Phi_0/2\pi \xi(0)^2$ and the current density with $j_0=\sigma_n\hbar/2e\tau_{GL}(0)\xi(0)$. Time is scaled in units of the
Ginzburg-Landau relaxation time $\tau_{GL}(0)=\pi \hbar/8k_BT_cu$, the electrostatic potential ($\varphi$) is in units of
$\varphi_0=\hbar/2e\tau_{GL}(0)$ ($\sigma_n $ is the normal-state conductivity, and $D$ is the diffusion constant). The parameter $u$ is equal to $5.79$
in accordance with Ref. \cite{Kramer} and we used $\gamma=40$. We put ${\bf A}=(Hx,0,0)$ in Eqs. (1,2) because we limit ourselves to the case when the
effect of the self-induced magnetic field is negligible. This is valid in the experimental situation because the width of the sample is much less than
the characteristic length $\Lambda=\lambda(0)^2/d_f$ ($d_f$ is the thickness of the sample).

Strictly speaking Eq. (1) is valid only very close to the critical temperature (see estimates for different low-temperature superconductors in Ref.
\cite{Watts-Tobin}). For example for bulk 'clean' Al the validity of Eq. (1) was derived only for the range $\Delta$T $\sim 10^{-4}$K near T$_c$. However
our Al samples are in the 'dirty' limit due to the small value of the mean path length $\ell$. As follows from Refs. \cite{Romijn,Sols} the relation
between current density $j$, the absolute value of the order parameter $|\psi|$ and the momentum $p$ are quite close to the Ginzburg-Landau relation
$j=p(1-p^2)=|\psi|^2p$ even when T$\to$ 0 for such samples. Besides, when we turn on the magnetic field and/or the transport current the density of
states of quasi-particles differs from the Bardeen-Schriffer-Cooper dependence \cite{Parks} and can become gapless \cite{Parks,Anthore} for high enough
magnetic fields and/or transport currents. In this case Eq. (1) should be valid at any temperature $T<T_c$ because they were actually derived in the
gapless limit (with $\gamma=0$) \cite{Schmid} or for small value of the gap $\Delta(T)\ll k_BT_c$ \cite{Kramer}.

The actual value of $\gamma$ for Al should be about $10^3$ because the time $\tau_E\sim 10^{-8}$s is quite large in this material. However, the use of
such a large $\gamma$-value is important if we intend to compare {\it quantitatively} the theoretical and experimental values for the critical current.
As will be shown below there exist two critical currents which we call first $I_{c1}$ and second $I_{c2}$ critical currents. The first critical current
is the current at which the sample goes to the non-resistive state and its value strongly depends on the value of the parameter $\gamma$ (in case of
strong heat dissipation - see text below). The second critical current has the meaning of the current at which the superconducting state becomes unstable
and it could be determined from a stability analysis of the stationary Ginzburg-Landau equations. It implies that the current $I_{c2}$ does not depend on
the $\gamma$ value which is the main reason why we are able to find quantitative agreement between theory and experiment for the position of the cusps in
the $I_{c2}(H)$ dependence (see below).

In our theoretical model we considered the geometry depicted in Fig. 4 which simulates the real experimental samples. To simplify our model we used
linear contacts instead of the wedge shaped contacts used in the experiment. The main difference between them is that for wedge shaped contacts the order
parameter is more suppressed in point A at low magnetic fields because the current density in this region is maximal (see Fig. 1). In order to inject the
current in our system we used normal metal-superconductor boundary conditions at the end of the leads, i.e. $\psi=0$ and $-\nabla \varphi=j$. At the
other boundaries we used the usual insulator-superconductor boundary conditions: $(i\nabla-{\bf A})\psi|_n=0$ and $\nabla \varphi|_n=0$.

We also took into account the change of the local temperature in the sample in the resistive state by adding the heat diffusion equation to Eqs. (1,2)
\begin{eqnarray} C_{eff}\frac{\partial T}{\partial t} & = &
K_{eff}\Delta T+d_f j_n^2/\sigma_n-h(T-T_0),
\end{eqnarray}
where $T_0$ is the bath temperature, $C_{eff}=(D_s C_s+d_f C_f)$ is the effective heat capacity, $K_{eff}=(D_s k_s+d_f k_f)$ is the effective heat
conductivity coefficient, and the heat transfer coefficient $h=k_s/D_s$ governs the heat removal from the sample. Here we used a model for the
temperature distribution in thin superconducting films discussed in details in Ref. \cite{Mintz} and $C_s$, $C_f$, $k_s$, $k_f$ are the heat capacity and
the heat conductivity of the substrate (subscript s) and film/sample (subscript f), respectively. In this model it is assumed that the thickness of the
substrate and the film $D_s+d_f$ is much smaller than the healing length $\Lambda_h=\sqrt{K_{eff}/h} \gg D_s+d_f$.

If heat removal is strong enough (large value for the coefficient $h$) we can neglect the effects due to the local change of the temperature. In the
opposite case the results will quantitatively depend on the ratio between the healing length and the sample parameters (width of the square and the ratio
between the value of the current density in the contacts and in the square). We chose our parameters in such a way that it optimizes the calculation time
(small value of $C_{eff}$) and we considered cases of large, intermediate and small value of the coefficient $h$. In dimensionless units [the same as Eq.
(1)] Eq. (3) may be written as follows
\begin{eqnarray} \widetilde{C_{eff}}\frac{\partial T}{\partial t} & = &
\widetilde{K_{eff}}\Delta T+j_n^2-\widetilde{h}(T-T_0),
\end{eqnarray}
where $\widetilde{C_{eff}}=(D_s C_s/d_f+C_f)T_c\sigma_n/\tau_{GL}(0)j_0^2$, $\widetilde{K_{eff}}=(D_s k_s/d_f+k_f)T_c\sigma_n/\xi^2(0)j_0^2$,
$\widetilde{h}=hT_c\sigma_n/d_fj_0^2$ and the temperature is measured in units of $T_c$. If $D_s C_s/d_f \ll C_f$ and $D_s k_s/d_f \ll k_f$ we can use
the Wiedemann-Franz law as an estimate for $C_f$ and $k_f$ and we obtain for $\widetilde{C_{eff}}=\pi^3/48 \simeq 0.65$ and
$\widetilde{K_{eff}}=\pi^4/48u^2 \simeq 0.06$ at a temperature close to $T_c$. These values should be considered only as a very rough estimate for the
real magnitudes because normally the following inequalities are valid: $D_s C_s/d_f \gg C_f$ and $D_s k_s/d_f \gg k_f$. Because of the uncertainty in the
actual values of $C_s$ and $k_s$ we used the following values: $\widetilde{C_{eff}}=0.03$ (to optimize calculation time), $\widetilde{K_{eff}}=0.06$, and
${\widetilde h}=2\cdot 10^{-3}$ (which corresponds almost to full heat removal at $T=0.9$), ${\widetilde h}=2\cdot 10^{-4}$(intermediate heat removal)
and ${\widetilde h}=2\cdot 10^{-5}$ (weak heat removal) \cite{ours} and a bath temperature of $T_0=0.9$. We checked that our results only weakly depend
on our choice of $\widetilde{C_{eff}}$ and $\widetilde{K_{eff}}$. As a boundary condition to Eq. (4) we take $\nabla T|_n=0$ which means that heat is
mainly transferred to the substrate. Only at the boundary between the normal metal and the superconductor we used boundary conditions with fixed
temperature $T_{NS}=T_0$. The healing length is equal to $\Lambda_h \sim 11 \xi(T=0.9T_c)$ for ${\widetilde h}=2 \cdot 10^{-5}$ and is comparable to the
size of the sample.

\begin{figure}[hbtp]
\includegraphics[width=0.45\textwidth]{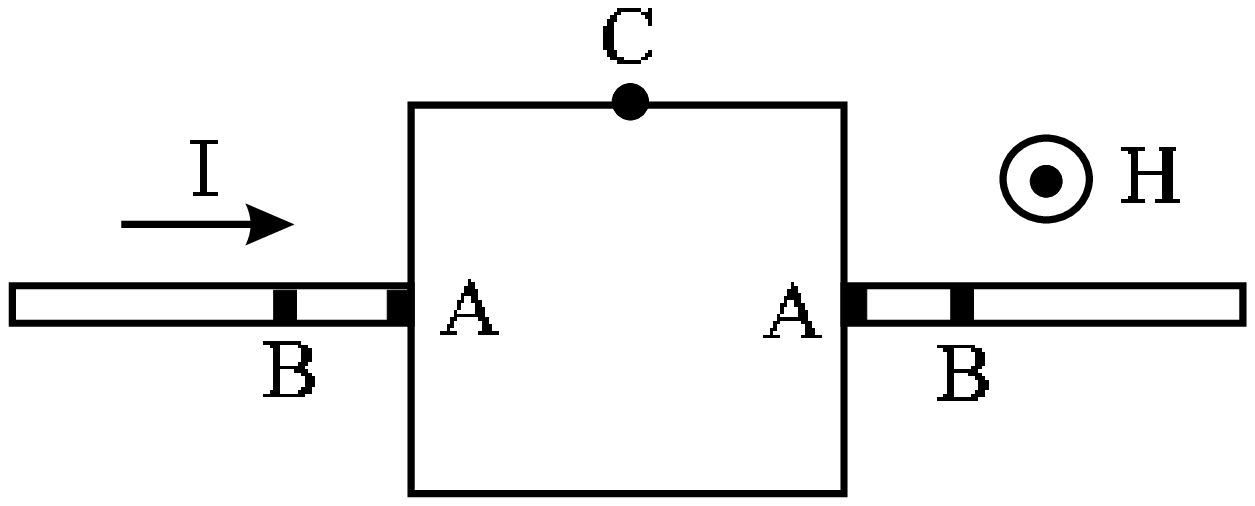}
\caption{Model geometry. Between points A and B phase slip centers appear at H$\lesssim$ H$_{c2}$. In point C we trace out the dependence of $|\psi|$ on
the magnetic field (see text below and Fig. 5).}
\end{figure}

Although in our numerical calculations we used units normalized at T=0 we will discuss here mainly the situation for a bath temperature of $T_0=0.9T_c$.
Therefore, it is more convenient to express the different quantities in units normalized at $T=T_0$. For example, under $H_{c2}$ we mean
$H_{c2}$($T=0.9T_c$).

First we studied the behavior of our model geometry sample in an applied magnetic field with zero transport current. In Fig. 5 we present the dependence
of the free energy of the square and the value of the order parameter in the center of the edge (i.e. at point C in Fig. 4) as a function of the magnetic
field. It turned out that for the chosen parameters (width of the square is 6$\xi$ and the size of the wires is 24$\xi \times \xi$ which are close to
experimental values with $\xi\simeq 0.333 \mu$m and H$_{c2} \simeq$ 2.95 mT) no single quantum vortex state exists in the square and only surface
superconductivity nucleates, i.e. the giant vortex state is present at H$>$0.8H$_{c2}$. Superconductivity vanishes in the superconducting square at
$H\gtrsim 2.71 H_{c2}$ which is much larger than the third critical field $H_{c3}=1.69 H_{c2}$ of a flat infinite surface, which is a consequence of the
shape and finite size of our system. By this value the vorticity in the system is equal to 10 and superconductivity may survive in the contacts up to
much higher values of the magnetic field.

\begin{figure}[hbtp]
\includegraphics[width=0.48\textwidth]{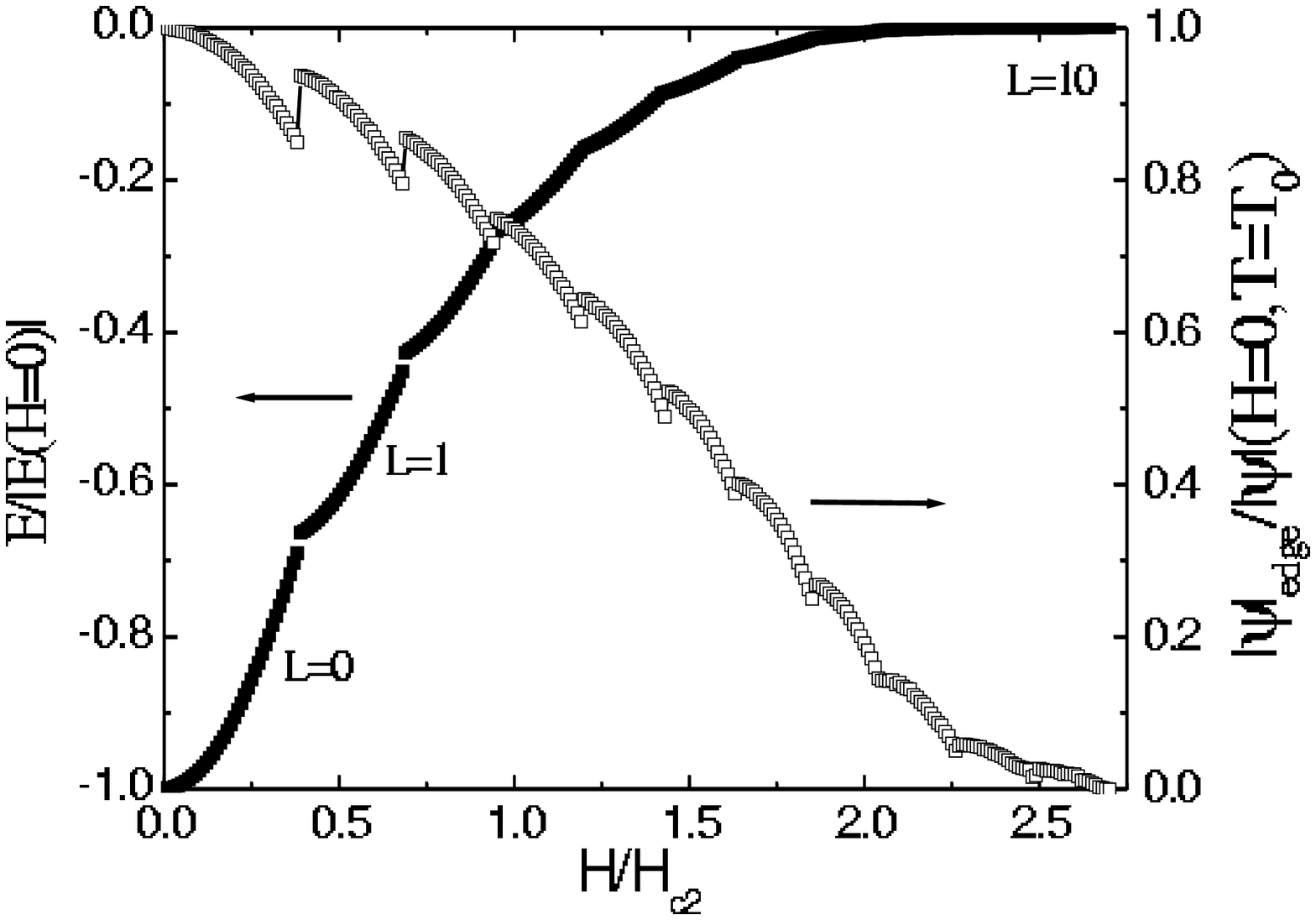}
\caption{Calculated dependence of the free energy (of the ground state) and the order parameter in the center of the edge of the superconducting square.}
\end{figure}

What will occur when we switch the transport current on? Let us consider first the situation when the heat removal is quite good and the temperature of
the sample is equal to the bath temperature. Then up to some magnetic field $H^* (H^* \simeq 0.92 H_{c2}$ for our parameters) the transition to the
resistive state from the superconducting state occurs via the appearance of phase slip centers in the contacts [see Figs. 6(a,b)]. Current $I_{c2}$ at
which this occurs slightly depends on the applied field because the order parameter in the square depends on $H$ (see Fig. 5) due to the induced
screening currents. Via proximity effects the variation in the order parameter in the square influences the order parameter in the contacts and hence the
critical current for nucleation of the phase slip centers.
\begin{figure}[hbtp]
\includegraphics[width=0.43\textwidth]{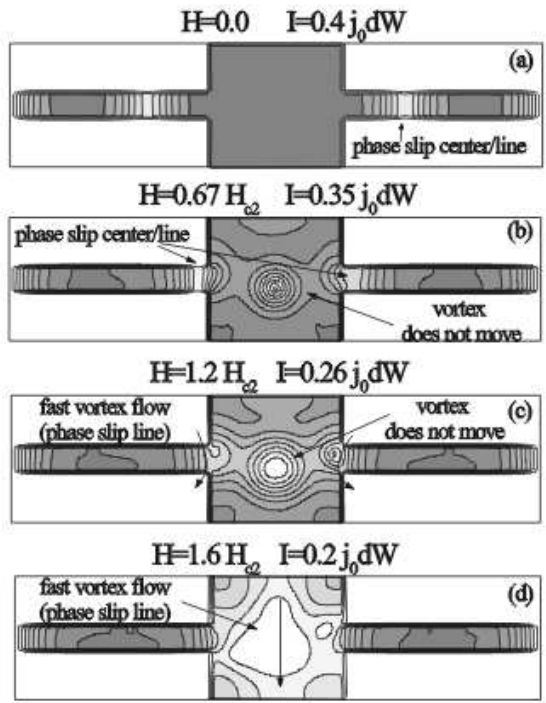}
\caption{Snap-shots of the order parameter distribution (dark color corresponds to the maximal value of $\Psi$ and grey to the minimal one) in a
superconducting square with attached leads at different magnetic fields and $I>I_{c2}$(H).}
\end{figure}

At higher magnetic fields the vortex flow regime starts at $I=I_{c2}(H)$. At fields close to $H^*$ this regime switches to the phase slip process near
the contacts [see Fig. 6(c)] while at fields higher than $H^*$ with further increase of the current it switches to the phase slip line regime in the
square [or more exactly a line along which the vortices move very fast \cite{Vodolazov2} - see Fig. 6(d)]. It occurs because the order parameter is
strongly suppressed in the square by the magnetic field while in the contacts its influence is less pronounced due to the small width of the contacts.
The voltage exhibits a jump at the transition from slow flux flow to the phase slip line/center regime. The larger $H$ the smaller this jump in the
voltage \cite{Vodolazov2}. The actual value of $H^*$, at which this change in the mechanism of destruction of superconducting state at $I=I_{c2}(H)$
occurs, depends on the width of the contact. The narrower the contacts the higher the field at which flux flow in the square starts before the phase slip
process occurs in the contacts.

We should note here that the position of the phase slip line in the contacts depends on the applied magnetic field. When the order parameter at the edge
of the square decreases, the phase slip center approaches the square and vice versa. For the case of wedge contacts we do not expect such a behavior
because in this case the order parameter is more suppressed around the narrowest point where the current density is maximal.
\begin{figure}[hbtp]
\includegraphics[width=0.45\textwidth]{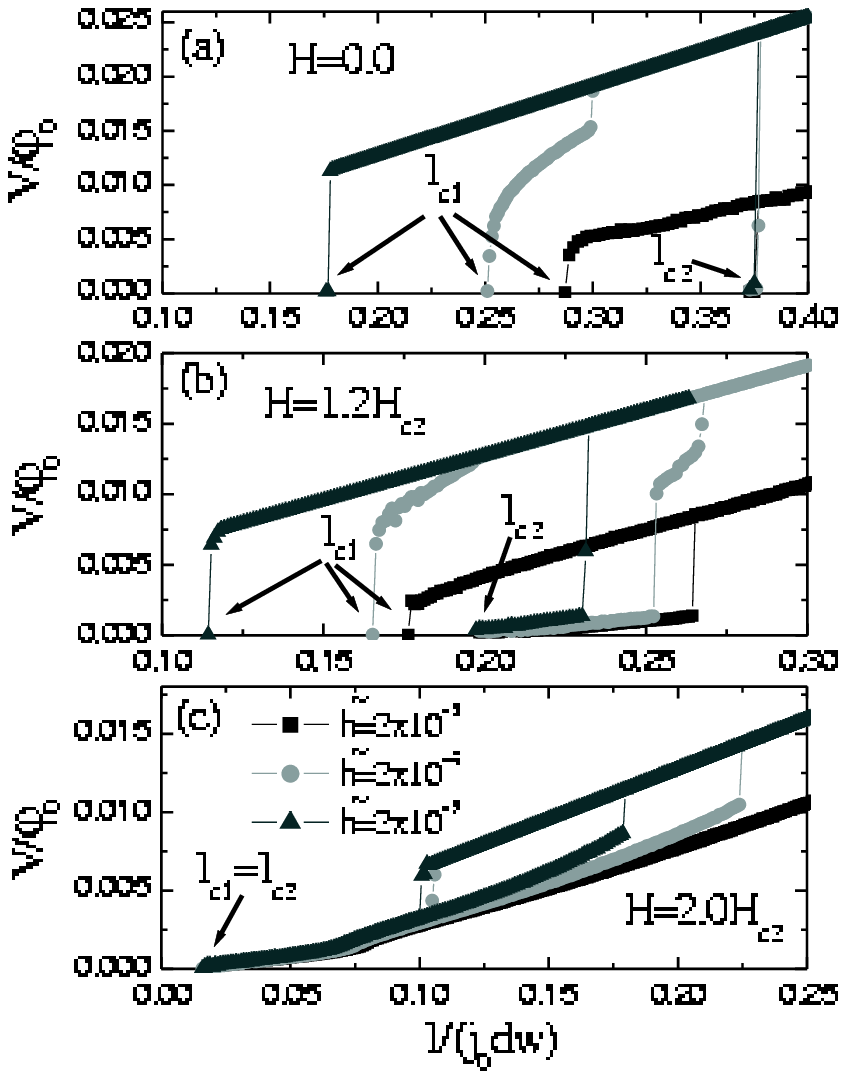}
\caption{Current-voltage characteristics of the superconducting square with contacts calculated for different heat removal coefficients $\widetilde{h}$
and magnetic fields. Current is normalized by the value $j_0dw$, where $d$ and $w$ are the lead thickness and width, respectively.}
\end{figure}

In Fig. 7 we present the I-V characteristics calculated for different values of the magnetic field. For low magnetic fields and large heat transfer
coefficient $\widetilde{h}$ the hysteresis in the I-V characteristics is an internal property of the phase slip process \cite{Ivlev,Michotte,Vodolazov2}.
At the current $I_{c1}(H$) the sample goes to the superconducting non-resistive state from the phase slip regime at fields less than some critical value
$H^{**}$ or from the flux flow regime for higher fields. The actual value of the field $H^{**}$ depends on the value of the coefficient $\gamma$ and the
width of the contact (see the above discussion for field $H^*$). With increasing $\gamma$, the minimal current at which the phase slip process is still
possible, decreases and hence the field $H^{**}$ increases, because the current at which the flux flow starts in the sample does not depend on the
relaxation times of the superconducting condensate.

The magnetic field effects the hysteresis through the local suppression of the order parameter \cite{Michotte}. This is the main origin of the decreasing
and finally the disappearance of the hysteresis at the transition from the superconducting state or slow vortex flow regime to the phase slip regime at
high magnetic fields. Another effect of the magnetic field is the slow increase of $I_{c1}$ at low magnetic fields. The reason for this is the same as
was found in Ref. \cite{Vodolazov2} - the nonuniform current density distribution in the contacts due to the applied magnetic field.
\begin{figure}[hbtp]
\includegraphics[width=0.45\textwidth]{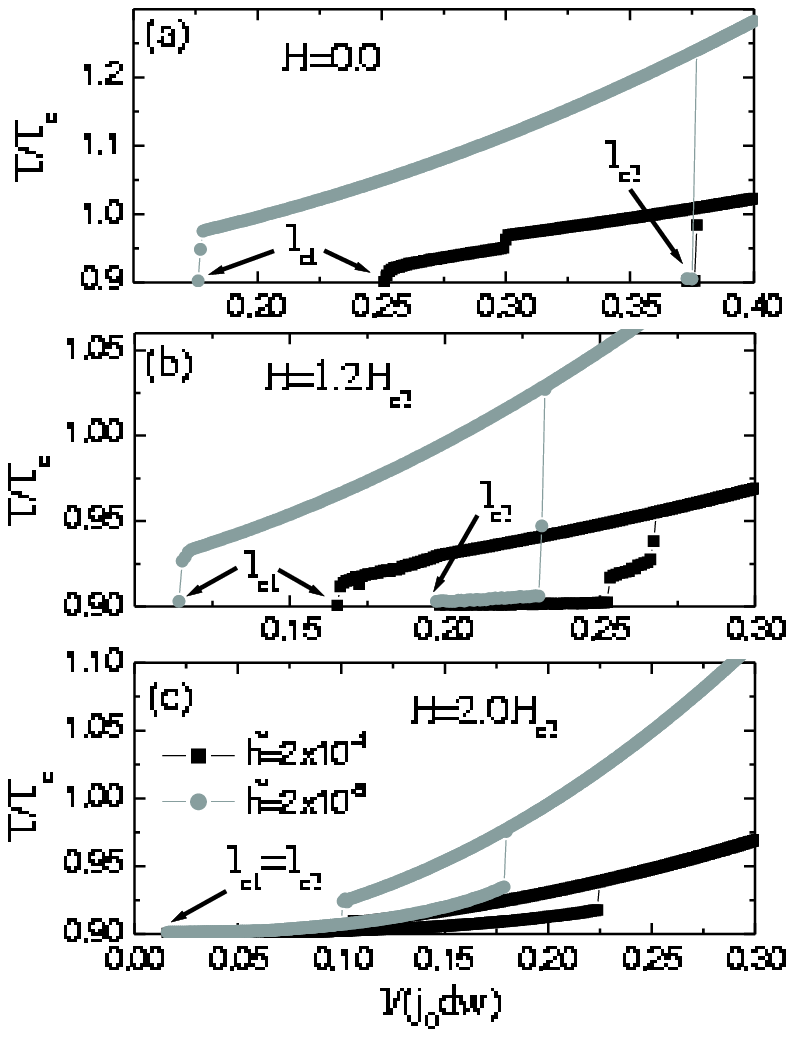}
\caption{Calculated dependence of the temperature of the superconducting square at different currents and magnetic fields. Results for
$\widetilde{h}=2\cdot 10^{-3}$ are not presented due to quite small changes (less than 1$\%$) in the temperature.}
\end{figure}

When the heat removal is not effective, then the I-V characteristics have a different shape at low and intermediate magnetic fields. In Fig. 7 we present
our results for two relatively small values of the heat transfer coefficient $\widetilde{h}$. At low and intermediate magnetic fields the whole sample
goes into the normal state (with $T> T_c$ or $T<T_c$) at the current $I=I_{c2}(H)$ because of the large heat dissipation which is connected with the
phase slip process. The value of that critical current is the same for any heat transfer coefficient $\widetilde{h}$ due to the absence of heat
dissipation in the "pure" superconducting state.

When we decrease the current, the temperature in the sample can become less than $T_c$ (see Fig. 8) while the sample will not go into the superconducting
state because at this temperature the current in the sample is too high in order that superconductivity can sustain it. Only when the temperature in the
sample becomes less than some critical temperature $T^*<T_c$ (which depends on the value of the heat transfer coefficient $\widetilde{h}$) then
superconductivity starts to nucleate in the square in places where the current density is minimal (near the corners of the square). For this current,
magnetic field, bath temperature and in the absence of local heating, the flux flow or phase slip processes are impossible (see Fig. 7 for high value of
$\widetilde h $) and the sample should go to the superconducting state.  But due to heat dissipation the actual temperature of the sample is still larger
than $T_0$ (see Fig. 8). Consequently the sample is in the resistive state with a resistance less than the normal one. The range of currents for which
such a process is possible depends on many parameters. For example, it increases with increasing magnetic field and decreasing heat transfer coefficient.

At high magnetic fields the critical currents $I_{c2}$ and $I_{c1}$ are quite small and even in the case of weak heat removal the I-V characteristics
resemble the ones with strong heat removal at low currents [Fig. 7(c)] due to weak heating [see Fig. 8(c)].
\begin{figure}[hbtp]
\includegraphics[width=0.48\textwidth]{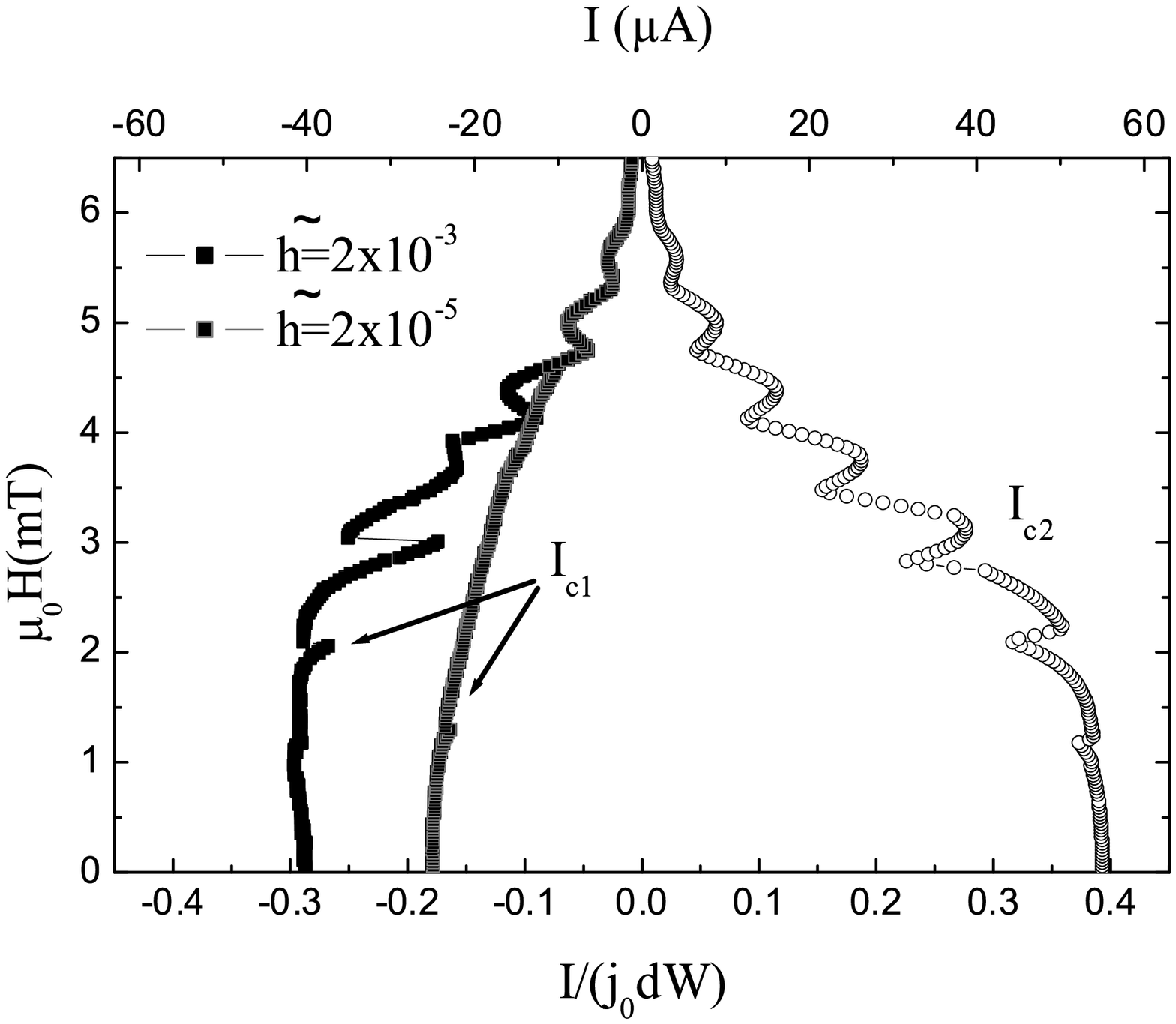}
\caption{Calculated dependence of the critical currents $I_{c2}$ and $I_{c1}$ (for $\widetilde{h}=2\cdot 10^{-3}$ and $\widetilde{h}=2\cdot 10^{-5}$) on
the applied magnetic field. $I_{c2}$ and $I_{c1}$ coincide for $H\gtrsim$3.9 mT (1.32$H_{c2}$) at strong heat transfer ($\widetilde{h}=2\cdot 10^{-3}$)
and for $H\gtrsim$4.6 mT (1.56 $H_ {c2}$) at weak heat transfer ($\widetilde{h}=2\cdot 10^{-5}$). There exists a nonmonotonous behavior both in the
$I_{c2}(H)$ and $I_{c1}(H)$ dependencies which is connected with a change in the vorticity in the superconducting square.}
\end{figure}

And finally in Fig. 9 we present the dependence of the currents $I_{c2}$ and $I_{c1}$ on the applied magnetic field for strong and weak heat removal.
There is a good quantitative agreement between the position of the cusps in the experimental and the theoretical dependencies of $I_{c2}(H)$. We explain
these cusps by abrupt changes in the vorticity and hence sharp changes in the order parameter distribution (see Fig. 5). The main difference between
theory and experiment is in the amplitude of the variation of $I_{c2}$ with $H$ (the theory predicts larger values) and in the value of this critical
current. We believe that this disagreement originates mainly from the difference in the real shape of the attached contacts (Fig. 1) and the contacts
used in our model (Fig. 4). In the experimental case there is a voltage lead which 'strengthen' the superconducting property near the narrowest point (as
it was shown in the quasi-one dimensional limit in Ref. \cite{Fink}) and actually shifts the position of the phase slip center further from the square.
Therefore, the effect of the variation of the order parameter in the square (with applied field) should be less pronounced on the phase slip process in
the contacts. Variations in $I_{c2}$ become stronger at $H\gtrsim H_{c2}$ when the resistive regime starts from the vortex flow regime in the square and
hence the effect of a change in the vorticity is more 'visible'.

We interpret the black right line in Fig. 3 as the line corresponding to the transition from the vortex flow to the phase slip line regime induced by
increasing current at which a jump in the heat dissipation occurs. Because the heat removal is not effective, the square will become  normal. Otherwise,
it will be in the superconducting resistive state with a resistance close to the normal one. Here we would like to stress the following. In aluminium the
decay length of the charge imbalance $\Lambda_Q$ or, in other words, the region where the normal current density is finite near the phase slip center
\cite{Skocpol} is quite large \cite{Klapwijk} $\Lambda_Q \simeq 50 \mu m$. The size of our sample is much less than $\Lambda_Q$. So in this case it is
quite difficult to distinguish between the normal and the superconducting resistive state because the differential resistance would be the same and equal
to the normal one (see Ref. [30] in Ref.\cite{Michotte}).

With decreasing current (in absolute value) we cross the left black line (Fig. 3) and at low magnetic fields almost immediately enter the zero resistance
state. Actually, the current $I_{c1}$ even increases a little. The same increase of $I_{c1}$ can be reproduced theoretically if we assume that these
transitions occur due to the decay of the phase slip process in the contacts at strong heat removal (see Fig. 9).

A comparison with the experiment also shows that with increasing $H$ we cross the left black line and do not enter the zero resistance state. Thus, we
are probably in the vortex flow state but with the sample temperature larger than the bath temperature. The sample can be in the resistive state up to
lower currents than it can be at the bath temperature (see Figs. 7,8 for small heat transfer coefficient).

At very high fields dissipation is not very important at currents close to the critical ones $I=I_{c2}(H)=I_{c1}(H)$ (due to their small value) and
besides there is no 'internal' hysteresis due to the phase slip process. As a result there is no hysteresis in the current when the non-zero resistance
state appears and when a fast change in the resistance occurs in our sample.

\section{Conclusion}

Hysteresis in the current-voltage characteristics of superconducting wires, films or mesoscopic samples may appear due to heat dissipation or/and due to
'internal' hysteresis connected with the existence of phase slip lines/centers. We believe that in our measurements we have both types of hysteresis
which are responsible for the observed effects. At low magnetic fields the sample enters the normal state due to the appearance of the phase slip process
and a strong heat dissipation at the critical current $I_{c2}$. Because of a discrete change of the vorticity in the superconducting square the order
parameter changes abruptly at some values of H and it leads to cusps in the dependence of $I_{c2}(H)$. At higher fields, instead of a high dissipative
phase slip process, we have slow vortex motion and heat dissipation results in a weak effect on the I-V characteristics for $I\gtrsim I_{c2}(H)$.

When we decrease the current (at fixed value of the magnetic field) the sample goes first from the normal to the superconducting resistive state (left
black line in Fig. 3) and than slow vortex flow starts in the superconducting strip (at high magnetic fields) or the phase slip process in the contacts
(at low magnetic field). Because the temperature of the sample may be higher than the bath temperature, the current, at which the sample goes to the
nonresistive state, may be smaller than $I_{c2}$ even if the resistive state starts as a vortex flow. So actually we 'need' heat dissipation to explain
this effect.

One of the main results of our paper is that we show (by a self-consistent solution of the time-dependent Ginzburg-Landau equation and the heat diffusion
equation) that {\it heat dissipation does not necessarily lead to the destruction of superconductivity} as it was supposed in a recent paper
\cite{Tinkham2} (in order to explain recent experiments on MoGe films (see references therein)). In these samples the value of the coefficient
$\widetilde h$ is quite small due to the small value of the normal conductivity and both critical currents $I_{c1}$ and $I_{c2}$ are rather large because
of the absence of the external magnetic field. In general, the situation may be more complicated when heating does not destroy superconductivity. In such
a case, heating leads to additional complexity in the dynamics of the superconducting condensate due to the local heating of the sample.

Another result is, that by comparing the experimental and theoretical $I_{c2}(H)$ dependence we may distinguish \cite{Schweigert} between the giant
vortex and the single quantum vortex state which appear in the sample. It allows, in principle, to study experimentally the transformation between these
two different configurations as a function of the shape, size of the sample and external magnetic field.

\begin{acknowledgements}

This work was supported by the Belgian Science Policy, GOA (University of Antwerp), the Research Fund K.U. Leuven GOA/2004/02, the ESF-programme VORTEX,
and the Flemish Science Foundation (FWO-Vl). M.M. acknowledges support from the Institute for the Promotion of Innovation through Science and Technology
in Flanders (IWT-Vlaanderen).

\end{acknowledgements}


\begin{references}
\bibitem{Gor'kov} L.P. Gor'kov, Zh. Eksp. Teor. Fiz. {\bf 36},
1918 (1959) [Sov. Phys. JETP {\bf 9}, 1364 (1959)]. We use a value which is different from the Bardeen-Cooper-Schrieffer value for $\Delta_0$ because Eq.
(1) is only valid near $T_c$ where $Delta_{T}=\Delta_0(1-T/T_c)^{1/2}$. Therefore, it should be considered only as an effective parameter rather than the
actual value for the gap at $T=0$ $\Delta_{BCS}\simeq 1.74 k_BT_c$.
\bibitem{ours} Our estimate of the experimental value for $\widetilde h$
gives $\widetilde h \simeq 2\cdot 10^{-6}$ if we take the heat conductivity of our upper SiO$_2$ substrate layer $d=500 nm$ at T=1.2 K as $k=6\cdot
10^{-5} W/mK$. Actually the heat conductivity of the main Si substrate is two orders of magnitude larger but in this case the condition $\Lambda_h \gg
D_s+d_f$ is not satisfied. Therefore, in the experiment the value of $\widetilde h$ should be in the range $10^{-3}-10^{-5}$.
\bibitem{our2} Strictly speaking it is valid only if $u \gtrsim 1$.
If $u \lesssim 1$ the supercondicting resistive state (with appearance of phase slip centers) may exist at $j>j_{c2}$. But in this case the resistive
state will be unstable at $j<j_{c2}$ and we will not observe the hysteresis in the current-voltage characteristic.
\end{references}
\end{document}